# The Formation of Glycolonitrile (HOCH$_2$CN) from Reactions of C$^+$ with HCN and HNC on Icy Grain Mantles

David E. Woon

Department of Chemistry, University of Illinois at Urbana-Champaign, 600 S. Mathews Avenue, Urbana, IL 61801 USA; dewoon@illinois.edu

## Abstract

Quantum chemical cluster calculations show that reactions of C$^+$ with HCN or HNC embedded in the surface of an icy grain mantle can account for the formation of a recently detected molecule, glycolonitrile, which is considered to be an important precursor to ribonucleic compounds. Reactions of cations deposited on ice mantles with minimal kinetic energy have been found theoretically to result in previously unknown pathways to significant organic compounds in protostellar systems and the interstellar medium. In density functional theory cluster calculations involving up to 24H$_2$O, C$^+$ reacts consistently with HCN embedded in ice to yield the neutral HOCHNC radical with no barrier, along with H$_3$O$^+$ as a byproduct. If HOCHNC then reacts with H, three species can be formed: HOCH$_2$NC (isocyanomethanol), HOCH$_2$CN (glycolonitrile), and HOCHNCH. For the C$^+$ + HNC reaction on ice, the HOCHCN and H$_2$OCCN radicals form as intermediates, the first of which is another direct precursor to glycolonitrile via H addition. In addition to characterizing reaction pathways, predictions are provided of the vibrational and electronic spectra of the HCN and HNC starting clusters and the HOCHNC ice-bound intermediate.

*Unified Astronomy Thesaurus Concepts:* Astrochemistry (75); Pre-biotic astrochemistry (2079); Protoplanetary disks (1300); Molecule formation (2076); Molecular spectroscopy (2095); Theoretical models (2107); Grain surface reactions

## 1. Introduction

The identification of previously unknown complex organic molecules (COMs) in diverse astrophysical environments continues to be of great interest for characterizing the present-day inventories of the interstellar and circumstellar media as well as protostellar disks and nebulae. Particularly for the latter, present day discoveries provide some of the best insight into what molecules may have been present in our own protosolar nebula and may thus have contributed directly or indirectly to the inventory of molecules available for prebiotic chemistry on Earth. In recent years, detections of molecules in protostellar nebulae and disks have emerged as important subsets of all deep space detections, with the lists of detections growing for protostellar disks such as TW Hydrae and DM Tauri and for the low-mass, solar-type protostar IRAS 16293-2422 B. For the latter, recent detections include $CH_3Cl$ (Fayolle et al. 2017), trans-HONO (Coutens et al. 2019), $HOCH_2CN$ or glycolonitrile (Zeng et al. 2019), and, most recently, propanal and propylene (Manigand et al. 2020a). In addition to detecting new species, it is of great interest to try to account for how they form. The focus of the present work is to describe the plausible formation of glycolonitrile via cation-molecule reactions occurring on icy grain mantles. Glycolonitrile is considered to be a precursor to ribonucleotides and is thus of considerable astrobiological interest (Ritson & Sutherland 2012, 2013).

The constraints on the formation of molecules in interstellar and protostellar systems make it challenging to identify viable pathways that can occur at very low temperatures and densities. It was recognized early in the study of astrochemistry that gas phase ion-molecule reactions may contribute many viable pathways because the reactions are often barrierless if they are exothermic (Watson 1973; Herbst & Klemperer 1973; see also Larsson et al. 2012). Later it was recognized that some neutral-neutral gas phase reactions also have no barriers and can contribute to interstellar chemistry (Chastaing et al. 1998; Kaiser 2002; Smith et al. 2004). Further enhancement of rate coefficients may occur for reactions between two neutral species on icy grain surfaces either because the energetics are altered favorably by the ice or because encounter rates are increased by confining reactants to a surface (Cuppen et al. 2017). It's also well known that ice mixtures can be chemically altered by energetic irradiation with photons (mostly UV but also x-ray) or cosmic rays, forming molecules at least as complex as amino acids (Bernstein et al. 2002, Muñoz Caro et al. 2002, Hudson et al. 2008). To round out the tableau of reactions that can contribute to interstellar and protostellar disk chemistry, we have found (Woon 2011, 2015) that reactions can occur when cations are deposited on an ice surface even at very low deposition energies, i.e., with almost no kinetic energy, which is the case for gas phase species in cold sources. [Radiolysis also involves charged particles, but energies are enormously larger: the distribution of galactic cosmic rays measured in the Solar System peaks around 100–500 MeV (Mewaldt et al. 2010)]. Quantum chemical cluster calculations have shown that low deposition energy cation-ice reactions may produce different products than the corresponding gas phase cation-neutral processes due the presence and active participation of water in the ice. Cation-ice reactions offer alternate pathways to commonly observed species, such as formic acid (HCOOH) from $HCO^+$ reacting with $H_2O$ in ice, methanol ($CH_3OH$) from $CH_3^+$ reacting with $H_2O$ in ice, and $CO_2$ from $OH^+$ reacting with CO embedded on ice (Woon 2011).

In a previous study (Woon 2015) reactions between $C^+$ and pure ice clusters were found to produce two products in about an equal ratio, HOC and $CO^-$. HOC is a precursor to methanol (by adding three H atoms to the C), while $CO^-$ is only stable due to interactions with $H_2O$ molecules in the ice. The logical next step in the study of $C^+$ ice reactions is to add impurities on the surface of pure water clusters for the cation to react with. In the present work, we found that reactions between $C^+$ ions and both HCN and HNC yield intermediates that are precursors to glycolonitrile. For the most part, the intermediates that form only need to react with one additional H atom to form glycolonitrile or its isocyano isomer. No reaction barriers were found to inhibit the addition of $C^+$ to HCN or HNC to form radical intermediates. Both HCN and HNC were considered since the abundance of HNC is comparable to that of HCN in protostellar systems (Markwick et al. 2002; Graninger et al. 2014).

In the remainder of this study, we will describe the methodology that was employed to model ice chemistry and spectra and then present the results of our investigation of the formation of glycolonitrile and other products from the reactions of $C^+$ with HCN and HNC, followed by the addition of H atoms. Before proceeding to the study of the title reactions, we will briefly describe the expansion of our library of water clusters up to 32 water molecules and the construction of initial base clusters of HCN and HNC embedded in water clusters. In addition to characterizing reaction pathways, we report vibrational and electronic spectra of the base clusters and one of the intermediate compounds that is formed in the ice after $C^+$ is deposited.

## 2. Methodology and Base Clusters

Quantum chemical calculations have been used for modeling ice chemistry for something over two decades (Zamirri et al. 2019). The majority of the calculations described below were performed with Gaussian 09 (Frisch et al. 2009). We have found (Woon 2015) that optimizations performed with density functional theory (DFT) at the B3LYP level (Becke 1993; Lee et al.1988; Stephens et al. 1994) followed by frequency calculations adequately account for vibrational frequencies (when scaled by 0.9824) as well as zero-point energy (ZPE) corrections. For optimizations, aug-cc-pVDZ correlation consistent basis sets are used for all atoms except hydrogen, where the cc-pVDZ set is used (Dunning 1989; Kendall et al. 1992). This level of theory is designated B3LYP/MVDZ (for mixed valence double zeta). Our prior work (Woon 2015) found that this level of theory acceptably accounts for IR frequencies of water and small water clusters, the water dimer binding energy, and long-range effects, even though it does not include dispersion corrections. Geometries, total energies, and ZPEs of species optimized in this work may be accessed at https://arxiv.org/abs/2010.xxxxx.

In the prior study (Woon 2015), this same level of theory (B3LYP/MVDZ) was used to predict electronic spectra of pure water clusters and clusters in which chemistry had occurred via time-dependent DFT (TDDFT) calculations (Bauernschmitt & Ahlrichs 1996; Casida et al. 1998; Stratmann et al. 1998). In that study, the degree of accuracy for describing electronic absorptions in amorphous ice left something to be desired. We have recently determined through an extensive set of benchmark calculations (Woon, unpublished) that the accuracy of predicting electronic spectra via TDDFT is considerably improved if we use the MPW1K functional (Lynch et al. 2000), as long as the basis set on hydrogen is upgraded to aug-cc-pVDZ. These calculations are designated TDDFT-MPW1K/AVDZ. All TDDFT calculations were performed using B3LYP/MVDZ geometries; a benchmark calculation in a modestly-sized $8H_2O$ cluster found that this introduces only small differences compared to a cluster reoptimized at the MPW1K/AVDZ level. 300 electronic states were calculated for each structure and used to generate a synthetic electronic absorption spectrum. Excitation energies of gas phase species used for references were calculated with multireference configuration interaction (MRCI+Q level) using an extended multiconfigurational self-consistent field (MCSCF) wavefunction, as calculated with MOLPRO (Werner et al. 2010).

The utility of spectroscopic predictions for ice chemistry studies is twofold. First, spectroscopy makes it possible to track change during reactions. Baseline behavior evolves as species are added or as they are chemically transformed. The second application of spectroscopy is the possibility of definitively identifying a species on the basis of one or more well-defined, consistent features, whether in the IR or UV-vis regions or both. Here, a target molecule (HCN or HNC) is added to pure water ice, and we can assess if the impurities have any features that stand out from the vibrational or electronic spectra of amorphous ice. Once the $C^+$ cation is introduced and new intermediates and other products form, we can assess what features have appeared or disappeared. All of these steps confirm that a reaction has occurred and hopefully offer features that would allow identification to be made if a commensurate experiment was performed.

In prior work (Woon 2015) we described the development of a library of water clusters ranging from 16–19 H$_2$O to model chemistry on amorphous ice. We have extended this library to include five larger clusters with 21, 25, 27, and 32 (×2 cases) water molecules. Figure 1(a,b) depicts the vibrational and electronic spectra of these clusters, comparing the spectra summed over the six smaller clusters against the spectra summed over the extended library of clusters (all eleven for the vibrational spectrum and the first nine for the electronic spectrum; the 32H$_2$O clusters are too large for use in subsequent open shell calculations and were excluded). See Woon (2015) for more details and discussion about the summation process. It is evident that the two sets of summed spectra do not differ that much. For the vibrational spectra depicted in Figure 1(a), both summed spectra reproduce two of the distinctive features of amorphous solid water (ASW) (Devlin et al. 2001) quite well: the broad intermolecular librational band that peaks just below 800 cm$^{-1}$ and the sharper, molecular water bending band near 1600 cm$^{-1}$; the dotted orange lines mark the centers of the 800 cm$^{-1}$ and 1670 cm$^{-1}$ features reported by Devlin et al. for unannealed ASW deposited at 15 K. Figure 1(b) compares the results at the B3LYP/MVDZ and MPW1K/AVDZ levels for the electronic spectrum of amorphous ice. The latter is shifted substantially with respected to the former, by about 0.5 eV to higher absorption energies. At the MPW1K/AVDZ level, the characteristic initial absorption peak occurs close to 8.4 eV, which is now in reasonable agreement with the experimental work of Kobayashi (1983); the dotted orange line marks the location the first peaks for absorption by both crystalline and amorphous ice at 8.6 eV.

Rather than deposit HCN or HNC directly on one of the pure amorphous water ice clusters—where the interactions are fairly weak—HCN base clusters were created by replacing a water molecule at the edge of one of the pure water base clusters with HCN and then optimizing the structure. Five clusters containing HCN and $18H_2O$, $20H_2O$ (×2), $24H_2O$, and $26H_2O$ were built in this manner. Three of these clusters (with $18H_2O$, $20H_2O$, and $24H_2O$) were converted to HNC-ice clusters by swapping C and N and optimizing to a new minimum. This strategy results in starting clusters that partially embed the target molecule among water molecules more realistically. Figure 2(a,b) depicts the summed vibrational and electronic spectra of the HCN and HNC base clusters. Included in each plot are the summed spectra from Figure 1(a,b) for pure water (SUMMED11 for vibrational spectra, SUMMED9 for electronic absorption spectra). For the vibrational spectra depicted in Figure 2(a), the water bending peaks were scaled to match in intensity. For HCN, a weak peak near 2080 cm$^{-1}$ stands out (weakly) against the pure water baseline. This is comparable to the feature at 2097 cm$^{-1}$ for HCN diluted 1:18 in water ice reported by Noble et al. (2013) and the feature at 2092 cm$^{-1}$ for HCN diluted 1:5 in water ice reported by Gerakines et al. (2004). For HNC, features around 1900–2000 cm$^{-1}$ stand out with greater intensity but with more variation. In both cases, the features are due to the CN stretch. For HCN the spread in the position of the feature is about 50 cm$^{-1}$. The spread is more substantial for HNC. For the electronic spectra shown in Figure 2(b), there is very little difference between the three cases. The first vertical excitation energies for gas phase HCN and HNC both exceed 8 eV [8.1 eV for HCN, 8.6 eV for HNC at the MRCI+Q/aug-cc-pVTZ level]; their excitations are evidently buried beneath the contributions from water.

### 3. $C^+$ + HCN reactions on ice

Five optimizations were performed where $C^+$ was allowed to react with HCN embedded in the five clusters described above. The ion was placed from 2.1–2.6 Å from the N of HCN. Figure 3(a) depicts a representative starting geometry. The reaction occurs on a $^2A$ surface, since the base clusters are all singlets and the ground state of $C^+$ is $^2P$. In each of the five cases, the same reaction sequence occurs and yields the same products. The transient species $HCNC^+$ forms first, but it immediately attacks a neighboring $H_2O$ and incorporates it to form $H_2OCHNC^+$. This species is also transient and quickly loses a proton from the O atom, leaving the neutral HOCHNC radical (the hydroxy isocyano methyl radical). The proton that departed transfers to one of the neighboring water molecules to form hydronium ($H_3O^+$), but it migrates from water to water through the cluster until a minimum is found. Figure 3(b) depicts a representative case of the intermediate products. The overall reaction is

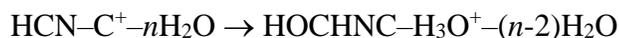
$$HCN–C^+–nH_2O \rightarrow HOCHNC–H_3O^+–(n\text{-}2)H_2O$$

Hence, two water molecules are consumed: one contributes the OH to intermediate HOCHNC, and another hosts the departing proton as hydronium. This entire process has no activation barriers whatsoever. It proceeds spontaneously from the initial starting structure to the one depicted in Figure 3(b). The same outcome was found to occur in all five cases that were run, which is rather atypical behavior (compare against the multiple outcomes in the reaction of $C^+$ and $H_2O$ on ice reviewed in the Introduction and the reaction of between $C^+$ and HNC on ice described below). Figures 3(c) and 3(d) respectively depict the structures obtained when H is added to HOCHNC to form $HOCH_2NC$ (isocyanomethanol) and the glycolonitrile isomer formed when the CN is flipped around. The transition state was located for this but is not depicted (see the gas phase structures shown in Figure 6). The energetics for the net reaction for a representative case will be discussed below, where a comparison between gas phase and ice surface behavior is considered.

The summed vibrational and electronic spectra of the five clusters where HOCHNC and $H_3O^+$ form from the reaction of $C^+$ with HCN embedded in ice are depicted in Figures 4 and 5, respectively. The summed spectra of the water clusters with unreacted HCN are included to depict the extent to which the clusters are altered by the chemistry. In both cases, the change is marked.

For the IR spectra in Figure 4, the intensities of the water bending feature near 1600 $cm^{-1}$ are scaled so they match. New features are clearly present on either side of this strong feature, while one feature present in the unreacted clusters essentially vanishes. This is the CN stretch described above. There is still motion associated with the CN stretch around 1930 $cm^{-1}$, but it becomes extremely weak in HOCHNC. The new features in the clusters where reactions have occurred are almost entirely due to motions involving $H_3O^+$, which are enhanced by the inherent charge of the species. Motion due to CH wagging contributes somewhat to the features between the water bending and librational features, but it is not especially strong or located at a consistent frequency.

For the UV-vis spectra in Figure 5, the water absorption features near 8.45 eV are scaled so they match. In the clusters where reactions have occurred, one of the new features that appears is the absorption between 6.5 and 7.0 eV, before the threshold is surpassed for water to absorb near 8 eV. It is likely that this absorption is due to HOCHNC, since $H_3O^+$ only begins to absorb around 12.0 eV in the gas phase (result from a MRCI+Q/aug-cc-pVTZ calculation). The other new feature, a strong absorption near 9 eV, may also be due to HOCHNC.

It is evident that if HCN were to be deposited on ice in a laboratory experiment, reactions with $C^+$ could be monitored with either vibrational or electronic absorption spectroscopy. We shall now consider an energetic analysis of forming glycolonitrile from the HOCHNC intermediate.

Once HOCHNC is formed on an ice mantle in the interstellar medium, it could react with other species, particularly radicals. One of the most likely radicals that could react with HOCHNC (or any other surface-bound radical, for that matter) would be H atoms, which are both abundant and very mobile on ice, even at 10-20 K (Al-Halabi & van Dishoeck 2007). It is assumed that H adatoms find and add to the radical intermediates (here and in the section on $C^+$ + HNC reactions on ice) without significant barriers. As Al-Halabi & van Dishoeck (2007) note, H atoms can travel for some distance (~30 Å) before they are thermalized by ice, which means they are overcoming diffusion barriers. Furthermore, H could also add to the radicals directly during deposition. The only condition is the usual one that the spins of the unpaired electrons on the radical intermediate and the H atom are antiparallel so that a covalent bond can form. The H atom was thus added directly to the radical without investigating the dissociative region of the interaction potential. Figure 6 depicts the energy relationships between the H + HOCHNC asymptote and several accessible products, including the transition state between isomers. Results are compared for the gas phase and a representative example from the five ice surface cases (the 18H$_2$O case). As a single case, the energetics should be treated as indicative rather than definitive (and one case is not enough to justify reporting vibrational or electronic spectra). The energy differences for addition and isomerization are found to be comparable between the gas phase and when the reaction occurs on ice. The ice has little effect on the energetics, which differ by no more than 3 kcal mol$^{-1}$ except for the HOCHNCH adduct, where the difference is about 5 kcal mol$^{-1}$.

If H adds to the central C atom of HOCHNC, HOCH$_2$NC (isocyanomethanol) is formed. It is an isomer of glycolonitrile, as noted above. If the CN of HOCH$_2$NC is flipped around, glycolonitrile is obtained. The energetics are clearly very favorable for all of this: even the transition state is substantially submerged, by nearly 50 kcal mol$^{-1}$ below the asymptote, so there is ample energy to drive the isomerization (if it is not absorbed by the ice; see below).

There are three additional points that merit discussion. (i) While isocyanomethanol is the most stable species formed by adding H to HOCHNC, there is a second adduct that can form without a barrier, HOCHNCH. While it is an odd structure, it is stable by over 60 kcal mol$^{-1}$ with respect to the asymptote in the case of the ice surface energetics. This pathway arises because the radical character of HOCHNC is *not* confined to the central C atom but is divided between both C atoms. The realization that this alternate formation pathway might be present was suggested by similar behavior observed in an analogous system in a prior study involving vinyl cyanide (Krim et al. 2019). (ii) We also investigated if there is sufficient reaction energy to eject HOCH$_2$CN from the surface. As shown in Figure 6, only about 14 kcal mol$^{-1}$ is needed to remove both glycolonitrile and a H$_2$O coordinated to it from the cluster shown in the Figure 3(d). There is plenty of reaction energy to do this. In principle, HOCH$_2$NC could also be ejected when formed, prior to isomerization, as well as HOCHNCH. (iii) However, investigating the probability that any species would actually be ejected from the surface would require a rigorous dynamics treatment of the system, which is beyond the scope of this work. It's also possible that the excess reaction energy is absorbed efficiently by the manifold of vibrational states, especially in a real ice. Several recent studies (Fredon et al. 2017, Fredon & Cuppen 2018, Pantaleone et al. 2020) concluded (along with previous studies cited therein) that desorption of molecules formed on ices is inefficient but not necessarily entirely precluded. Outcomes depend greatly on the specific reaction. In summation, the pure energetics indicate that at least three different compounds could form from the addition of H to the HOCHNC intermediate, and one or all of them could be ejected from the ice. The branching ratios for this depend on how well energy is absorbed by the ice: Efficient absorption would favor stabilization of HOCH$_2$NC and HOCHNCH, trapped in the ice; less efficient absorption would favor the formation of glycolonitrile and perhaps its ejection into the gas phase. The dispersal of energy in ice-bound reactions is an open issue where more theoretical and experimental work is needed. The conclusions drawn here reflect these uncertainties.

## 4. C$^+$ + HNC reactions on ice

Three examples of the reaction of C$^+$ with HNC on ice were run using the clusters described above. Without going into as much detail as described above for HCN, glycolonitrile can also be formed from the $C^+$ + HNC reaction, but the sequence is somewhat different and the outcome less definitive. As in the HCN reaction, the $C^+$ first adds to HNC to form transient $HNCC^+$. This species quickly loses a proton and forms neutral doublet CCN. CCN then attacks water to form $H_2OCCN$ in all three examples. In just one case, $H_2OCCN$ interacts with ice to effect H transfer from O to C to yield HOCHCN. The net reaction for this one outcome is thus

$$HNC\text{–}C^+\text{–}nH_2O \rightarrow HOCHCN\text{–}H_3O^+\text{–}(n\text{-}2)H_2O$$

in analogy to the reaction of $C^+$ with HCN. While HOCHCN is a direct precursor to glycolonitrile, this pathway would seem to yield the compound less reliably, since HNC is less abundant than HCN and the reaction of $C^+$ with HNC only yields the direct precursor in one of three cases considered. However, the reaction of $C^+$ with HNC has been shown by this work to yield a direct precursor to glycolonitrile in at least one case, and it may prove to be the most efficient pathway to its formation since HNC is nearly as abundant as HCN in some sources.

## 5. Plausibility of $C^+$ reactions with HCN/HNC in IRAS 16293-2422 B

While glycolonitrile may eventually be identified in other sources, at present it has only been detected in IRAS 16293-2422 B. [It was sought unsuccessfully toward Sgr B2(N) by Margulès et al. (2017).] The most promising pathway proposed above requires a number of species to be present: HCN, $C^+$, H atoms, and icy grain mantles. The presence of HCN in IRAS 16293-2422 B is well known and has been the subject of a high-resolution survey by Takakuwa et al. (2007). There has apparently not been a study or survey of $C^+$ toward IRAS 16293-2422 B, but the modeling study of Doty et al. (2004)

predicts $C^+$ to be present, albeit with an abundance about three orders of magnitude less than HCN. The study by Doty et al. predicts a copious amount of H atoms relative to HCN. Finally, various observational studies toward IRAS 16293-2422 B have inferred that ice chemistry may account for the presence of many complex organics in the source (e. g., Persson et al. 2018; Manigand et al. 2020a, 2020b). In fact, the components required to form glycolonitrile by cation-ion chemistry are comparably plentiful in other sources, so it may be widely distributed in protostellar systems if either of the proposed pathways account for most of its formation.

## 6. Conclusions & Acknowledgments

This work characterized the formation of glycolonitrile and other intriguing precursors to more complex organic molecules via $C^+$ reactions with HCN or HNC on icy grain mantles, where the cation is deposited with only the energy possessed by any cold species in the source. Predictions of both vibrational and electronic spectra indicate the changes that would be observable first when HCN or HNC is added to ice (in agreement with prior experiment in the case of HCN) and then subsequently when $C^+$ is deposited. In astrophysical sources where H atoms are also present, the HOCHNC intermediate that forms from the $C^+$ + HCN reaction is expected to yield $HOCH_2CN$, $HOCH_2NC$, and possibly HOCHNCH after hydrogenation; the HOCHCN intermediate that forms from the $C^+$ + HNC reaction (in one case) also yields $HOCH_2CN$ after hydrogenation. Product distributions will be heavily impacted by how efficiently the ice absorbs the copious energy produced by the reactions. The reaction of $C^+$ with HNC may in fact be the more favored pathway to glycolonitrile if either HOCHNC or $HOCH_2NC$ produced by $C^+$ + HCN reactions is stabilized before $HOCH_2CN$ can be formed through isomerization.

However, those intermediates are also of interest; finding them would support the pathways presented here. Astrophysical searches for radical HOCHNC and closed-shell HOCH$_2$NC would thus appear to be warranted if their rotational spectra should become available in the future. Experimental confirmation of any of the processes reported in the present study or the two previous ones (Woon 2011, 2015) would demonstrate that cation-ion reactions need to be included in modeling studies of interstellar and protostellar sources.

In this and the previous studies, H$_3$O$^+$ is often produced as a byproduct of cation-ice chemistry. If cation-ice reactions are important, then H$_3$O$^+$ may be a commonly present species in astrophysical ices. Hydronium evidently has pronounced vibrational features (as evident in Figure 4), so it may be observable. However, the protons formed from cation-ice reactions may not be predominantly associated with H$_2$O as H$_3$O$^+$ but with other species with larger proton affinities. The likely candidate to be a sink for protons in real astrophysical ice mixtures is ammonia, due to its abundance and large proton affinity. Ammonium (NH$_4^+$) may thus be one of the most common species found in astrophysical ices where cation-ice reactions are common.

The support of the NASA Emerging Worlds program through grant NNX16AM09G is gratefully acknowledged. Results from a collaboration with Dr. Lahouari Krim of Sorbonne Université were found to provide insight into the present work.

# Figures & Figure Captions

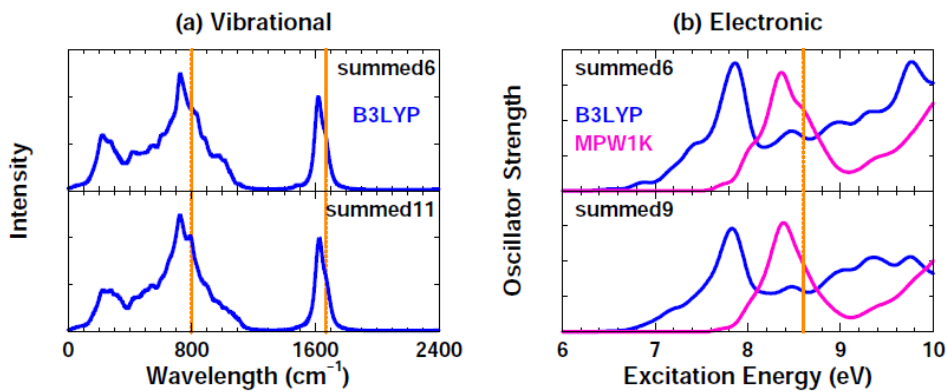

**Figure 1.** (a) Vibrational and (b) electronic spectra for water base clusters. Top panels on each side are summed over 6 clusters with 16–19 $H_2O$; the lower panel vibrational spectrum is summed over 11 clusters with 16–32 $H_2O$; the lower panel electronic spectra are summed over 9 clusters with 16–27 $H_2O$. For the electronic spectrum, results from the B3LYP and MPW1K functionals are compared. Experimental features are indicated with dotted orange lines (see text).

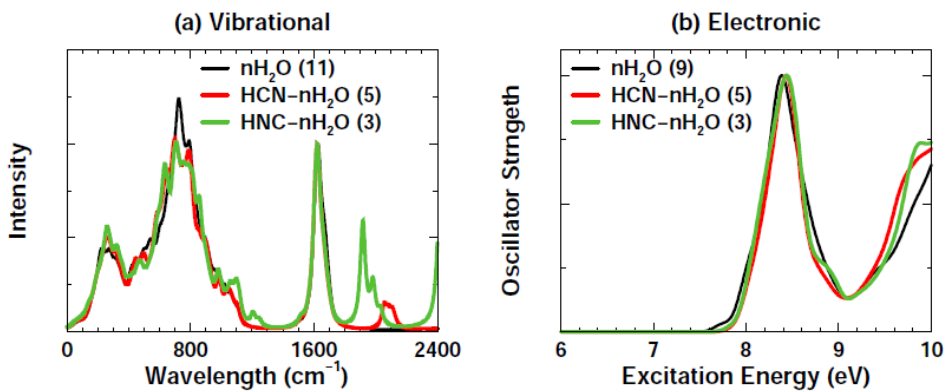

**Figure 2.** (a) Vibrational and (b) electronic spectra for HCN-$nH_2O$ and HNC-$nH_2O$ clusters compared against pure water base clusters. Values in parentheses indicate the number of individual spectra contributing to each summed spectrum.

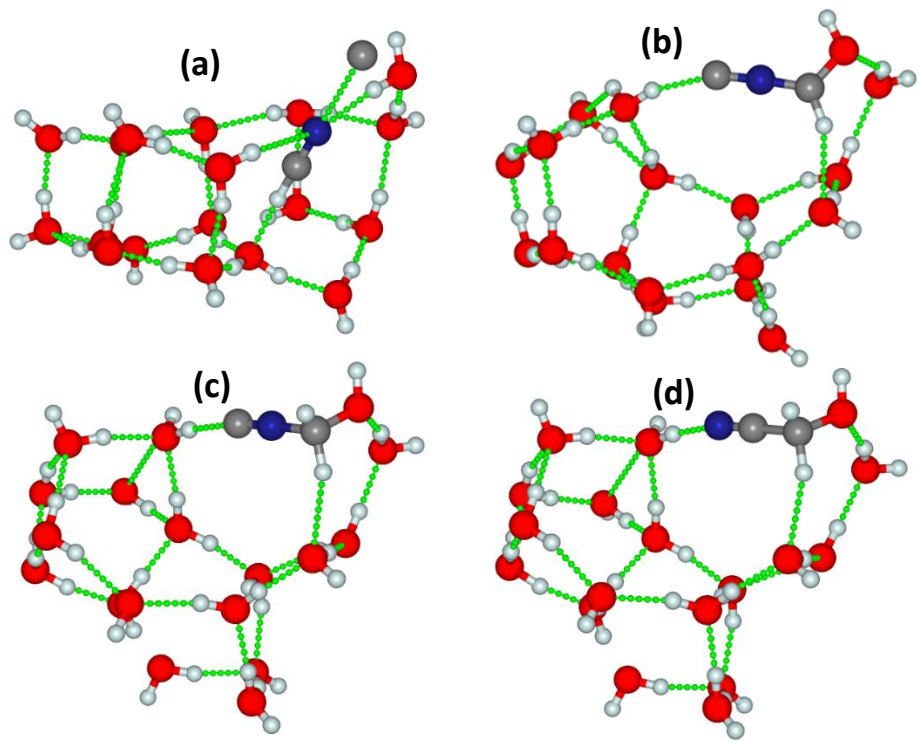

**Figure 3.** Representative structures, 18H$_2$O case (a) initial geometry: C$^+$ + HCN–$n$H$_2$O (b) intermediate: HOCHNC–H$_3$O$^+$–($n$-2)H$_2$O (c) H atom adduct: HOCH$_2$NC–H$_3$O$^+$–($n$-2)H$_2$O (d) glyconitrile isomer: HOCH$_2$CN–H$_3$O$^+$–($n$-2)H$_2$O

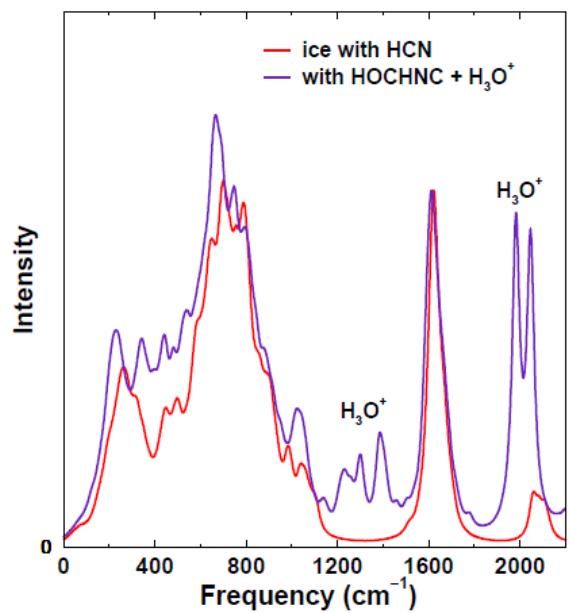

**Figure 4.** Summed vibrational spectrum of HOCHNC and $H_3O^+$ formed from the reaction of $C^+$ and HCN compared with the summed spectrum of 5 water clusters with unreacted HCN.

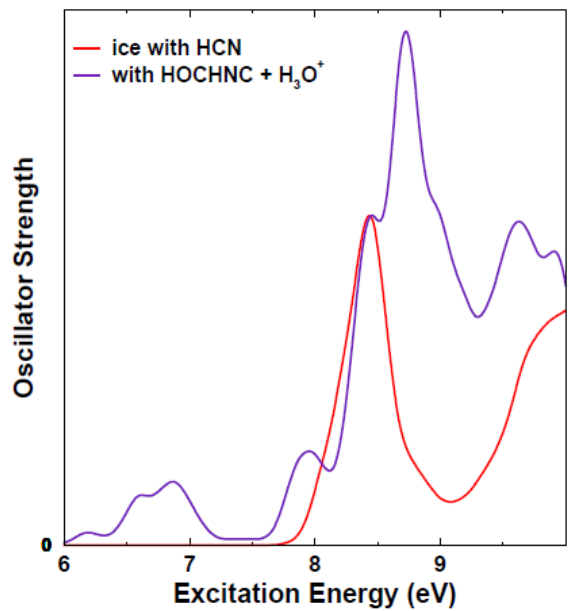

**Figure 5.** Summed electronic spectrum of HOCHNC and $H_3O^+$ formed from the reaction of $C^+$ and HCN compared with the summed spectrum of 5 water clusters with unreacted HCN.

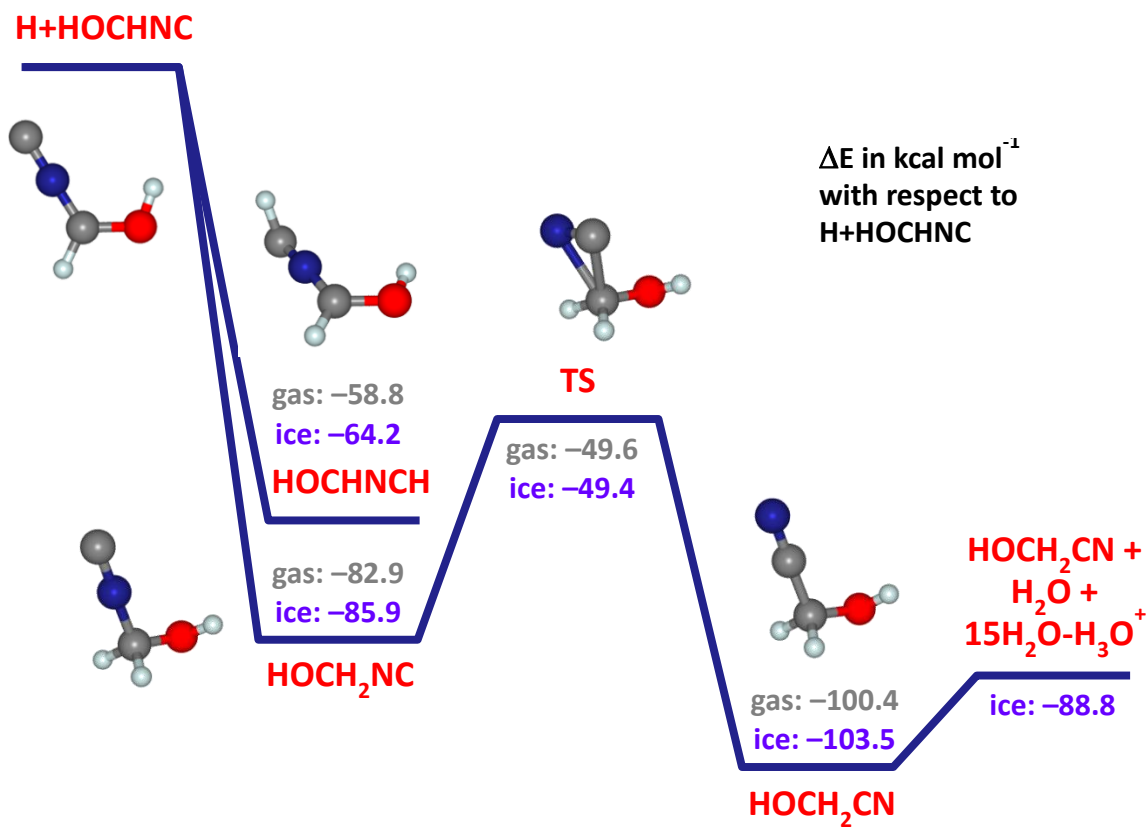

**Figure 6.** Potential energy diagram for adding H to HOCHNC (formed when $C^+$ reacts with HCN on ice). Energy differences in kcal mol$^{-1}$ with respect to H + HOCHNC are depicted for both gas phase and ice surface reactions. Isocyanomethanol (HOCH$_2$NC) forms first, but it may rearrange into the more stable isomer, glycolonitrile (HOCH$_2$CN), via the transition (TS) where the CN flips around. Also shown is the product asymptote for the ice surface case where glyconitrile and one of the water molecules are ejected from the ice. Structures for the gas phase reaction are depicted. HOCHNCH can also form in the initial H addition step.

Supplemental Information for "The Formation of Glycolonitrile (HOCH$_2$CN) from Reactions of C$^+$ with HCN and HNC on Icy Grain Mantles" by David E. Woon

1. Cartesian coordinates (Å), total energies (E$_h$), and zero-point energies (kcal mol$^{-1}$) for optimized product clusters of C$^+$ + HCN reactions (5 cases)

```
===========================================
HCN-18H2O + C+ -> (HOCHNC)-(H3O+)-16H2O (1)
===========================================
  58
EB3LYP = -1507.502901  ZPE = 300.914
  O      2.444275    -2.094001    -2.878865
  H      3.878963    -2.215882    -1.753859
  H      2.578331    -2.141005    -3.833610
  O      4.506573    -2.262076    -1.000016
  H      4.734824    -0.987858    -0.241830
  H      4.112170    -2.891746    -0.366078
  O      2.929477    -3.439779     1.088580
  H      2.079793    -3.376394     0.591929
  H      2.970661    -4.326770     1.467747
  O      0.813498    -2.911355    -0.545708
  H      1.042886    -3.177388    -1.450451
  H     -0.115897    -3.192590    -0.374272
  O     -5.372059    -2.659140     0.242553
  H     -6.043897    -1.084075    -1.142750
  H     -5.234931    -3.478428    -0.258200
  O     -3.418425     0.765285     2.023966
  H     -4.250705    -1.369096     1.450017
  H     -4.015424     1.258316     2.600121
  O     -3.849841     1.293289    -0.547202
  H     -3.000644     1.323416    -1.023112
  H     -3.673244     0.995190     1.088638
  O     -0.930814     1.820629     1.638050
  H     -1.752839     1.387293     1.970180
  H     -0.149092     1.428238     2.072357
  O     -0.615354     4.512233     1.160237
  H     -0.829615     3.608839     1.481873
  H     -1.167593     5.131448     1.650761
  O      1.721765     1.108487     2.423984
  H      2.013562     1.770429     3.065143
  H      2.113581     1.374270     1.558384
  O      2.804595     1.384524    -0.115216
  H      2.173738     0.815765    -0.625652
  H      2.820806     2.259844    -0.586257
  O      4.788142    -0.180663     0.411353
  H      4.119108     0.519392     0.104973
  H      4.392196    -0.530839     1.301667
  H      2.826541    -0.501433     2.645189
  O      3.588302    -1.085560     2.464865
  H      3.239832    -1.958102     2.188027
  H     -4.477059     0.753642    -1.066611
  O     -5.745195    -0.389748    -1.747620
  H     -6.537582    -0.047522    -2.179648
  H      1.954141    -1.265798    -2.695952
```

```
O     1.034443    -0.213007    -1.428511
H     0.228294     0.360087    -1.453596
H    -0.968541     1.633938    -0.148796
O    -1.028009     1.551241    -1.128689
H    -0.783359     2.439917    -1.455647
O    -0.105925     4.252802    -1.452838
H    -0.357505     4.505249    -0.527236
H    -0.518856     4.895007    -2.043201
O     2.601221     3.656943    -1.488481
H     1.681480     3.998261    -1.523802
H     3.190161     4.419966    -1.487172
H     0.794522    -1.026929    -0.942439
C    -4.209373    -2.238647     0.799422
N    -3.046041    -2.778905     0.447196
C    -1.970606    -3.217209     0.175456
```
============================================
HCN-20H2O + C+ -> (HOCHNC)-(H3O+)-18H2O (1)
============================================
```
  64
EB3LYP = -1660.425700  ZPE = 332.401
O    -1.365590    -1.062857    -1.034255
H     1.331696    -1.370302    -0.766445
H    -1.978679    -0.288417    -1.152877
O    -2.728357     1.247076    -1.440678
H    -3.578944     1.101514    -1.900746
H    -2.975151     1.768018    -0.645164
O    -5.431793     0.507975    -1.984619
H    -5.374400    -0.416709    -2.310060
H    -6.163711     0.918009    -2.463399
O    -3.595322    -1.588710     1.909332
H    -4.207033    -0.899903     1.533993
H    -4.156185    -2.244129     2.339990
O    -5.149946     0.361357     0.944686
H    -5.408607     0.358909     0.005703
H    -4.674632     1.200673     1.075552
O    -1.242457    -0.871973     2.839033
H    -2.152238    -1.149231     2.515463
H    -1.340683    -0.728691     3.789774
O    -3.290763     2.548666     0.960708
H    -2.429502     2.348614     1.398955
H    -3.467643     3.488207     1.091976
O    -0.814855     1.704852     1.733133
H    -0.885341     0.788129     2.077625
H    -0.441519     1.648622     0.829983
O     1.013024     3.660129     2.593355
H     0.321819     2.973037     2.528570
H     1.043505     3.960116     3.508888
O     1.610766    -0.388633    -0.960687
H     2.320441    -0.050562    -0.274620
H     0.854324     0.266850    -1.030762
O     0.012403     1.677489    -1.017075
H    -0.829254     1.710900    -1.509249
H     0.638113     2.338937    -1.395568
O     3.071181     2.942195     0.939594
H     2.788352     3.268003     0.067774
H     2.409701     3.266605     1.590660
O     2.173877     3.138230    -1.824663
```

```
H      2.224387     3.971569    -2.309231
H      2.813401     2.522400    -2.261422
O      3.931815     1.359624    -2.902757
H      4.327428     0.611038    -2.415945
H      3.928587     1.107244    -3.833327
O      5.145307    -0.642439    -1.235996
H      4.855722    -1.561753    -1.034679
H      6.110416    -0.659153    -1.263333
O      3.386946     0.390883     0.654949
H      3.304776     1.382442     0.842374
H      4.215055     0.250250     0.153755
O      3.449108    -2.064154     2.130151
H      3.440048    -1.110166     1.930179
H      2.532826    -2.288421     2.374418
O      0.639994    -2.800098     2.255477
H      0.395964    -3.604712     2.730297
H     -0.047615    -2.128328     2.485155
O      1.217696    -2.826023    -0.468203
H      0.950447    -2.937773     0.471052
H      2.137450    -3.155385    -0.525853
H      3.932976    -2.781917     0.686840
O      4.060892    -3.044648    -0.268385
H      4.576299    -3.861476    -0.270262
H     -1.290032    -3.095151    -1.473698
C     -1.900362    -2.200664    -1.512635
N     -3.134877    -2.204159    -2.016588
C     -4.242945    -2.046055    -2.430868
```
==========================================
HCN-20H2O + C+ -> (HOCHNC)-(H3O+)-18H2O (2)
==========================================
```
  64
EB3LYP = -1660.420991  ZPE = 330.524
O     -1.472949    -0.042571    -0.562094
H      0.251001     0.507643     0.064068
H     -1.611183    -0.822871     0.074849
O     -1.811628    -1.932607     1.208334
H     -2.352026    -1.483112     1.909601
H     -2.394365    -2.623845     0.836596
O     -3.423411    -0.486023     2.786653
H     -3.149635     0.477805     2.673528
H     -3.657972    -0.611943     3.713791
O     -2.821104     1.982270     2.230609
H     -1.894701     2.290810     2.099467
H     -3.292292     2.164760     1.399137
O     -4.371075     1.957487    -0.262786
H     -4.586458     0.996167    -0.146692
H     -5.217283     2.420173    -0.215714
O     -4.960117    -0.589529     0.406961
H     -4.604463    -0.658377     1.315079
H     -4.757300    -1.436542    -0.017338
O     -2.459843     2.832285    -2.038047
H     -3.176536     2.513739    -1.439609
H     -2.903212     3.202191    -2.811171
O     -3.730981    -3.387708    -0.288402
H     -3.307152    -3.588359    -1.140038
H     -4.250110    -4.171072    -0.062470
O      1.172662    -4.304010    -2.254779
```

```
H        0.258678     -4.008250     -2.429104
H        1.404449     -4.910271     -2.968750
O        1.123034      0.497257      0.500184
H        2.190666      0.100007     -0.513210
H        1.059552     -0.254541      1.160404
O        0.883875     -1.582785      2.118794
H        0.045914     -2.030084      1.909087
H        1.605863     -2.237143      2.063343
O        3.004755     -2.694164     -1.077580
H        3.090090     -2.996311     -0.147494
H        2.344255     -3.265454     -1.538582
O        3.294463     -2.998331      1.653645
H        3.623101     -3.801818      2.075653
H        3.952433     -2.286347      1.855318
O        4.998558     -0.939487      2.032795
H        5.101796     -0.247427      1.350021
H        5.153250     -0.516673      2.885380
O        5.090011      0.867505     -0.155894
H        4.826434      1.829882     -0.040696
H        5.921214      0.870640     -0.648408
O        2.915661     -0.231000     -1.165535
H        2.942009     -1.302343     -1.158462
H        3.794706      0.155831     -0.875956
O        2.158494      3.855870     -1.592756
H        2.336545      4.351548     -2.400324
H        1.223541      4.034699     -1.356573
O       -0.489410      4.259966     -0.811689
H       -0.788418      5.170757     -0.696868
H       -1.218510      3.786383     -1.285251
O       -0.258935      2.869050      1.702287
H       -0.288508      3.405814      0.888591
H        0.401637      2.180801      1.525271
H        3.471708      3.583852     -0.521320
O        4.267923      3.336396      0.008197
H        4.306293      3.926730      0.768526
H       -1.637808      0.398827     -2.583106
C       -1.568451     -0.397325     -1.848794
N       -1.533906     -1.680146     -2.205070
C       -1.491503     -2.855045     -2.425582
```
==========================================
HCN-24H2O + C+ -> (HOCHNC)-(H3O+)-22H2O (1)
==========================================
  76
EB3LYP = -1966.250551  ZPE = 396.732
```
O        1.099284      0.305823     -0.168888
H        0.517949      1.034830     -0.504546
H        1.983930      0.535122     -0.543947
O        3.337699      1.089391     -1.617822
H        4.167889      0.845159     -1.139998
H        3.284101      0.485020     -2.385765
O        5.308872      0.100063     -0.051367
H        4.997811      0.507111      0.812130
H        6.261544      0.247796     -0.094114
O        4.365821      1.331640      2.055916
H        3.919320      2.174628      1.816782
H        3.794648      0.911078      2.722763
O        2.298892      0.194933      3.754091
```

```
H     1.998175    -0.604712     3.263140
H     2.481111    -0.078138     4.661474
O     1.297617    -1.622576     1.957923
H     0.424584    -2.003636     2.209409
H     1.124693    -0.988789     1.236337
O     2.214652    -3.606298     0.146626
H     3.004819    -3.208363    -0.299899
H     2.015138    -3.009907     0.895466
O     4.253760    -2.360234    -1.120855
H     4.695385    -1.619989    -0.661908
H     3.963565    -1.999705    -1.975126
O     0.150611    -4.051047    -1.290022
H     1.000927    -3.905126    -0.727607
H     0.378343    -4.719988    -1.949442
O     2.908429    -1.031836    -3.380543
H     1.966798    -1.238282    -3.182772
H     3.037903    -1.144619    -4.329897
O     0.369882    -1.377920    -2.323656
H     0.172885    -2.302339    -2.069694
H     0.600698    -0.919352    -1.485763
O    -1.376593     0.728732    -3.043409
H    -0.867627    -0.104737    -2.986797
H    -1.663667     0.826930    -3.958485
O    -0.320117     2.399023    -1.168593
H    -0.738710     2.009940    -1.969351
H     0.425817     2.971533    -1.448970
O     2.183415     3.546424    -1.508512
H     2.731074     2.746061    -1.688589
H     2.512527     4.252826    -2.077746
O     2.847403     3.563687     1.371496
H     2.662545     3.730899     0.431189
H     1.991723     3.313990     1.758132
H     0.861604     1.322257     1.557431
O     0.582137     1.905243     2.286431
H     0.998176     1.500006     3.071938
O    -4.203976     3.704944    -1.525658
H    -3.954623     4.347273    -0.828718
H    -4.635546     4.215199    -2.219756
O    -3.194469     5.158343     0.603875
H    -3.624499     5.685778     1.286620
H    -2.685717     4.462832     1.057471
O    -1.778586     2.739861     1.109983
H    -1.301027     2.775221     0.245324
H    -1.068279     2.516266     1.752034
O    -3.365753     0.743022     0.730652
H    -3.597799    -0.757299     1.701650
H    -2.801224     1.571570     0.927717
O    -4.096405    -3.542196     0.350904
H    -4.351246    -2.991878    -0.476894
H    -3.190730    -4.030522     0.236750
O    -1.840729    -4.693971     0.306435
H    -1.839410    -5.652066     0.434248
H    -1.098962    -4.480149    -0.342627
O    -1.060990    -2.850243     2.563994
H    -1.229088    -3.609263     1.983769
H    -1.906017    -2.377074     2.622333
H    -4.008460    -2.856455     1.089708
```

```
  O    -3.722539   -1.638426    2.128415
  H    -4.351139   -1.507921    2.851571
  H    -4.071298    1.765839   -1.013019
  C    -3.933332    0.815516   -0.491440
  N    -4.308980   -0.355752   -1.004435
  C    -4.549048   -1.489791   -1.270597
==========================================
HCN-26H2O + C+ -> (HOCHNC)-(H3O+)-24H2O (1)
==========================================
  82
EB3LYP = -2119.155737  ZPE = 427.396
  O     1.114126    1.267877    0.239229
  H     0.530407    0.492679    0.111797
  H     1.727457    0.986000    0.956781
  O     2.511134    0.375593    2.452483
  H     3.435931    0.547907    2.173354
  H     2.257293    1.131288    3.020799
  O     4.770075    1.228327    1.087332
  H     4.804118    0.711931    0.245228
  H     5.687140    1.342239    1.366673
  O     4.782555   -0.268060   -1.195391
  H     5.193696   -1.146547   -1.142299
  H     3.985773   -0.391684   -1.734959
  O     2.137026   -0.702746   -2.845933
  H     1.907044    0.306080   -2.743840
  H     2.372847   -0.836464   -3.774218
  O     1.501843    1.715045   -2.433189
  H     0.554330    1.788295   -2.728616
  H     1.463197    1.713773   -1.451613
  O     1.588365    4.576645   -0.936720
  H     2.219822    4.283482   -0.236341
  H     1.788610    4.048243   -1.721078
  O     3.235726    3.742296    1.096723
  H     3.831310    2.978777    0.994184
  H     2.786219    3.603305    1.947640
  O    -0.916164    4.907435   -0.179886
  H     0.027436    4.826443   -0.510065
  H    -1.050538    5.842313    0.020155
  O     1.592874    2.866980    3.400926
  H     0.712923    2.984877    2.973900
  H     1.537891    3.265227    4.278163
  O    -0.574413    2.852722    1.694508
  H    -0.826978    3.667731    1.211929
  H     0.043975    2.391661    1.079439
  O    -2.110208    0.546066    1.665713
  H    -1.741981    1.455163    1.733943
  H    -3.061733    0.587388    1.872919
  O    -0.519762   -0.960306    0.129208
  H    -1.298979   -0.497350    0.532146
  H    -0.003055   -1.383538    0.872948
  O     1.083254   -2.005857    1.948009
  H     1.599712   -1.303558    2.393985
  H     0.582714   -2.528218    2.599904
  H    -0.214073   -1.604782   -1.124684
  O     0.203281   -2.065755   -1.975992
  H     1.011479   -1.508606   -2.363896
  O    -2.466355   -1.428617    3.787441
```

```
H    -2.090248   -2.277193    3.488680
H    -1.977681   -0.751355    3.290974
O    -1.038267   -3.677373    2.714766
H    -1.109467   -4.506940    3.202863
H    -1.266605   -3.879736    1.778268
O    -1.476428   -4.045164   -0.017103
H    -2.202469   -3.557958   -0.471185
H    -1.551816   -4.958482   -0.320249
O    -3.464584   -2.638802   -1.373469
H    -3.218046   -1.948570   -2.040072
H    -4.152358   -3.171150   -1.793926
O    -5.141686   -0.610273    0.034590
H    -5.089585   -0.554916    1.007924
H    -4.622081   -1.388112   -0.224150
O    -4.636043   -0.215405    2.754987
H    -4.029937   -0.808869    3.261125
H    -5.288463    0.127454    3.376939
O    -4.399382    1.212324   -1.835560
H    -4.728829    0.729513   -1.042187
H    -3.943570    2.021198   -1.549940
O    -2.708875    3.594476   -1.840397
H    -3.169617    4.237148   -2.394633
H    -2.126104    4.119982   -1.244520
O    -1.060895    1.730901   -3.163090
H    -1.650833    2.408430   -2.774391
H    -1.576985    0.905198   -3.180306
H    -3.529005    0.155140   -2.651055
O    -2.926122   -0.536429   -3.074716
H    -3.245397   -0.647028   -3.979857
O     1.717320   -4.393485   -1.293914
H     0.507697   -2.966098   -1.732803
H     2.248369   -4.782050   -2.007091
H     2.042786   -3.597098    0.615803
C     2.535170   -3.942022   -0.289826
N     3.805737   -3.644318   -0.557180
C     4.944653   -3.341254   -0.759283
```

2. Cartesian coordinates (Å), total energies ($E_h$), and zero-point energies (kcal mol$^{-1}$) for optimized product clusters of $C^+$ + HNC reactions (3 cases)

```
==========================================
HNC-18H2O + C+ -> (HOCHCN)-(H3O+)-16H2O (1)
==========================================
 58
EB3LYP = -1507.530109  ZPE = 297.502
 O    -3.160721   -1.571684    0.484189
 H    -3.752892   -2.248492   -1.410385
 H    -3.645657   -0.659652    0.195209
 O    -4.090941    0.516014   -0.388585
 H    -3.306374    0.972820   -0.782581
 H    -4.655435    1.162546    0.096973
 O    -5.748611    2.176297    0.948020
 H    -6.592947    2.434832    0.524879
 H    -5.467391    2.937088    1.468420
 O    -8.152306    2.943140   -0.220773
```

```
H         -8.992674       2.576731       0.081487
H         -8.246803       3.070557      -1.172910
O          1.162972      -3.485116      -2.036279
H          0.370008      -3.839696      -1.554458
H          1.291394      -4.064187      -2.798351
O          3.185940      -2.634784      -0.512934
H          2.461659      -3.010706      -1.086006
H          3.884273      -3.302438      -0.488499
O          2.454585      -1.716458       1.846903
H          2.386191      -0.701355       1.685253
H          2.682406      -2.156195       0.972554
O          3.841047       0.209043      -1.117889
H          3.772303      -0.759200      -1.183491
H          3.301808       0.597010      -1.847592
O          5.765319       1.810618       0.127622
H          5.265444       1.213109      -0.463380
H          6.450238       2.231758      -0.404373
O          2.026030       1.414533      -2.732699
H          2.130200       1.982603      -3.505472
H          1.542920       1.938034      -2.041218
O          0.744097       2.474160      -0.592918
H          0.980674       1.818217       0.085535
H          1.140609       3.310057      -0.255259
O         -1.791869       1.329651      -1.520673
H         -1.124184       1.925152      -1.141681
H         -1.281389       0.541586      -1.786890
H          0.806802      -0.111833      -2.625748
O          0.159195      -0.658117      -2.149029
H          0.395818      -1.583724      -2.316482
H          1.619403      -2.149531       2.314813
O          0.525319      -2.825294       2.942265
H          0.670280      -3.298333       3.769668
H         -2.320151      -1.681662       2.043110
O         -1.853067      -1.712151       2.906315
H         -0.395538      -2.435614       2.959741
H          2.939643       0.604997       0.320481
O          2.424711       0.725429       1.157909
H          2.915122       1.422235       1.648183
O          4.038635       2.867393       1.891930
H          4.775066       2.595990       1.286330
H          4.440625       3.049328       2.750414
O          2.236467       4.513740       0.445781
H          2.903852       4.126437       1.044785
H          2.027303       5.392685       0.781599
H         -2.528927      -1.883296       3.572607
N         -1.175840      -4.106563      -0.753332
C         -2.054472      -3.332152      -0.645786
C         -3.070061      -2.381720      -0.570413
============================================
HNC-20H2O + C+ -> (H2OCCN)-(H3O+)-18H2O (1)
============================================
  64
EB3LYP = -1660.390409  ZPE = 331.085
O         -1.818112       0.065010      -0.846084
H         -0.839044       0.080440      -0.686766
H         -2.202025      -0.247812       0.018848
O         -3.023126      -0.767290       1.435204
```

```
H    -3.438613    -0.001746     1.900485
H    -3.765765    -1.290443     1.086238
O    -4.312328     1.532329     2.234949
H    -3.773758     2.244192     1.737720
H    -4.544873     1.887671     3.101811
O    -3.236840     2.219951    -0.841895
H    -4.186431     1.787836    -0.672274
H    -2.580435     1.448595    -0.980666
O    -5.434065     1.175228    -0.260471
H    -5.406609     1.296982     0.713029
H    -5.463632     0.206420    -0.404097
O     1.769577     5.458386    -0.740414
H     0.905990     5.116295    -0.431038
H     1.682344     6.414643    -0.822451
O    -5.068489    -1.620169    -0.475350
H    -4.494018    -1.830581    -1.242217
H    -5.786893    -2.265398    -0.479683
O    -3.037343    -1.829571    -2.403488
H    -3.185330    -1.551070    -3.316746
H    -2.487410    -1.125409    -1.981692
O    -1.629971    -4.244800    -1.906920
H    -2.145170    -3.462312    -2.184664
H    -1.823869    -4.949702    -2.535811
O     0.752974    -0.141527     0.032759
H     1.611218    -1.103063    -0.761063
H     0.434135    -0.610718     0.854792
O    -0.303058    -1.497992     2.056196
H    -1.274813    -1.454084     2.057297
H    -0.037103    -2.432483     2.142073
O     0.841743    -3.950856    -0.857503
H     0.773776    -4.119760     0.108076
H    -0.050332    -4.077437    -1.261316
O     0.955146    -4.036542     1.901503
H     0.796268    -4.834836     2.420238
H     1.900292    -3.782669     2.060403
O     3.512179    -3.242066     2.133801
H     3.953470    -2.895781     1.332540
H     3.825992    -2.710599     2.874825
O     4.513592    -2.269594    -0.339131
H     5.088651    -1.448023    -0.267598
H     5.027855    -2.916230    -0.840119
O     2.079056    -1.856651    -1.289637
H     1.533446    -2.763266    -1.132809
H     3.024003    -1.973987    -0.974287
O     5.100102     2.087051    -1.420990
H     5.483214     2.467671    -2.218855
H     4.652873     2.815398    -0.936606
O     3.749127     3.901067     0.134687
H     4.237654     4.328558     0.848712
H     3.105041     4.575256    -0.206884
O     1.613853     2.176058     0.902385
H     2.447593     2.639755     0.693841
H     1.154106     0.729828     0.326636
H     5.667610     0.736752    -0.599204
O     5.895590    -0.080291    -0.087680
H     6.816630    -0.005145     0.184059
H     0.917489     2.846315     0.781653
```

```
 N     -0.560820     4.068922     0.184312
 C     -1.622201     3.551596     0.196314
 C     -2.875240     2.969345     0.388354
==========================================
HNC-24H2O + C+ -> (H2OCCN)-(H3O+)-20H2O (1)
==========================================
  76
EB3LYP = -1966.245425  ZPE = 396.103
 O     -0.672426    -0.448993     0.645159
 H     -0.846792     0.499491     0.834660
 H     -1.525509    -0.884484     0.877115
 O     -3.015770    -1.574387     1.623596
 H     -3.526126    -2.100190     0.957248
 H     -2.565762    -2.213910     2.206638
 O     -4.146021    -2.892956    -0.438711
 H     -4.133229    -2.161717    -1.129217
 H     -5.022427    -3.294423    -0.482736
 O     -4.163001    -0.844749    -2.060362
 H     -4.662666    -0.122476    -1.625577
 H     -3.387536    -0.437359    -2.488040
 O     -1.830167     0.444608    -3.168825
 H     -1.051537    -0.108833    -2.894951
 H     -1.830445     0.475191    -4.133691
 O      0.114649    -0.983220    -2.014309
 H      1.076025    -0.763550    -2.126333
 H     -0.117053    -0.750634    -1.091368
 O      0.205273    -3.794910    -1.693403
 H     -0.615098    -4.071938    -1.217811
 H      0.075559    -2.859783    -1.954700
 O     -1.866143    -4.502712    -0.094148
 H     -2.711044    -4.047252    -0.282366
 H     -1.660412    -4.307536     0.832930
 O      2.093760    -3.862741     0.003754
 H      1.336163    -3.884269    -0.696905
 H      2.144607    -4.753380     0.374587
 O     -1.112615    -3.399707     2.697162
 H     -0.243838    -2.958967     2.652190
 H     -1.107175    -3.924697     3.507758
 O      1.299559    -1.903967     1.768827
 H      1.548247    -2.612543     1.120744
 H      0.609811    -1.327958     1.338029
 O      3.368981    -0.494763     2.028803
 H      2.494304    -1.090056     1.982712
 H      4.147211    -0.935341     1.525016
 O     -1.829856     2.072229     1.016012
 H     -1.483537     2.788886     1.587681
 H     -2.683781     1.791929     1.406897
 O     -4.160113     0.859567     1.977795
 H     -3.828056    -0.069949     2.007233
 H     -4.618318     1.018305     2.812107
 O     -5.554408     0.952910    -0.511120
 H     -5.187296     1.054504     0.386420
 H     -6.025350     1.766304    -0.723896
 H     -1.801529     2.472128    -0.733083
 O     -1.620773     2.760909    -1.654691
 H     -1.704645     1.979568    -2.245782
 O     -0.553169     4.071099     2.535184
```

```
H    -0.144429    5.230962    1.265163
H     0.202955    3.466401    2.766849
O     0.005935    5.748528    0.441179
H     0.552167    6.510811    0.664612
H     0.176892    4.976793   -0.907211
O     0.219769    4.431515   -1.760542
H     1.157199    4.060237   -1.904316
H    -0.528335    3.690788   -1.756133
O     2.607835    3.632990   -2.161263
H     3.110048    2.913419   -1.697119
H     2.948410    3.684827   -3.061704
O     5.297673   -1.533003    0.755919
H     6.047283   -1.933327    1.211914
H     5.069771   -2.095028   -0.024500
O     4.246526   -2.890428   -1.309857
H     4.661197   -3.485917   -1.946451
H     3.493508   -3.379807   -0.892533
O     2.759845   -0.543194   -2.277813
H     3.314776   -1.321108   -2.100859
H     3.230128    0.218043   -1.879283
H     3.755188    1.387212    0.004406
O     3.975518    1.635142   -0.975753
H     4.935009    1.740169   -1.011701
H    -0.900218    4.416568    3.367058
N     1.427908    2.259149    2.773396
C     2.217765    1.534769    2.279419
C     3.192809    0.868848    1.541211
```

---

3. Cartesian coordinates (Å), total energies ($E_h$), and zero-point energies (kcal mol$^{-1}$) for structures on the H addition pathway (gas phase).

H + HOCHNC → HOCH$_2$NC → HOCH$_2$CN

```
==========================================
H (reactant)
==========================================
   1
EB3LYP =    -0.501258  ZPE =   0.000
 H    0.000000    0.000000    0.000000
==========================================
HOCHNC (reactant)
==========================================
   6
EB3LYP = -207.314373  ZPE =  22.548
 O    1.355600    0.763319    0.000000
 H    1.712116   -0.137136    0.000000
 C    0.000000    0.720654    0.000000
 H   -0.508818    1.677505    0.000000
 N   -0.658195   -0.437171    0.000000
 C   -1.240122   -1.485108    0.000000
==========================================
HOCH2NC (isocyanomethanol intermediate)
==========================================
   7
```

```
EB3LYP =  -207.961952  ZPE =   31.463
 H   -1.396975   -1.048546    0.634687
 O   -1.461919   -0.440918   -0.114297
 H   -0.658348    1.272649   -0.801729
 C   -0.513675    0.582495    0.038521
 H   -0.635230    1.130621    0.986294
 N    0.835982    0.092213   -0.003980
 C    1.936013   -0.327974   -0.018025
===========================================
TS (isomerization transition state)
===========================================
   7
EB3LYP =  -207.906716  ZPE =   30.135
 H   -1.518767   -1.075890    0.044743
 O   -1.408616   -0.184635   -0.317467
 C   -0.429287    0.466547    0.350988
 H   -0.320828    1.481712   -0.030440
 H   -0.452374    0.423850    1.442332
 C    1.086354   -0.643553    0.322863
 N    1.374071    0.244206   -0.422858
===========================================
HOCH2CN (glycolonotrile product)
===========================================
   7
EB3LYP =  -207.990067  ZPE =   31.611
 H   -1.394640   -1.095718    0.600616
 O   -1.517888   -0.450834   -0.108292
 H   -0.732891    1.284230   -0.798044
 C   -0.569149    0.592284    0.038813
 H   -0.709209    1.155726    0.976901
 C    0.827257    0.113241   -0.006614
 N    1.918742   -0.281531   -0.015191
```

4. Cartesian coordinates (Å), total energies ($E_h$), and zero-point energies (kcal mol$^{-1}$) for structures on the H addition pathway (ice).

H + HOCHNC–16H$_2$O–H$_3$O$^+$ → HOCH$_2$NC–16H$_2$O–H$_3$O$^+$ → HOCH$_2$CN–16H$_2$O–H$_3$O$^+$ → HOCH$_2$CN + H$_2$O + 15H$_2$O–H$_3$O$^+$ (desorption of H$_2$O and glycolonitrile from ice)

```
===========================================
H (reactant)
===========================================
   1
EB3LYP =    -0.501258  ZPE =    0.000
 H    0.000000    0.000000    0.000000
===========================================
HOCHNC-16H2O-H3O+ (reactant)
===========================================
  58
EB3LYP = -1507.502901  ZPE =  300.914
 O    2.444275   -2.094001   -2.878865
 H    3.878963   -2.215882   -1.753859
 H    2.578331   -2.141005   -3.833610
 O    4.506573   -2.262076   -1.000016
```

```
H         4.734824    -0.987858    -0.241830
H         4.112170    -2.891746    -0.366078
O         2.929477    -3.439779     1.088580
H         2.079793    -3.376394     0.591929
H         2.970661    -4.326770     1.467747
O         0.813498    -2.911355    -0.545708
H         1.042886    -3.177388    -1.450451
H        -0.115897    -3.192590    -0.374272
O        -5.372059    -2.659140     0.242553
H        -6.043897    -1.084075    -1.142750
H        -5.234931    -3.478428    -0.258200
O        -3.418425     0.765285     2.023966
H        -4.250705    -1.369096     1.450017
H        -4.015424     1.258316     2.600121
O        -3.849841     1.293289    -0.547202
H        -3.000644     1.323416    -1.023112
H        -3.673244     0.995190     1.088638
O        -0.930814     1.820629     1.638050
H        -1.752839     1.387293     1.970180
H        -0.149092     1.428238     2.072357
O        -0.615354     4.512233     1.160237
H        -0.829615     3.608839     1.481873
H        -1.167593     5.131448     1.650761
O         1.721765     1.108487     2.423984
H         2.013562     1.770429     3.065143
H         2.113581     1.374270     1.558384
O         2.804595     1.384524    -0.115216
H         2.173738     0.815765    -0.625652
H         2.820806     2.259844    -0.586257
O         4.788142    -0.180663     0.411353
H         4.119108     0.519392     0.104973
H         4.392196    -0.530839     1.301667
H         2.826541    -0.501433     2.645189
O         3.588302    -1.085560     2.464865
H         3.239832    -1.958102     2.188027
H        -4.477059     0.753642    -1.066611
O        -5.745195    -0.389748    -1.747620
H        -6.537582    -0.047522    -2.179648
H         1.954141    -1.265798    -2.695952
O         1.034443    -0.213007    -1.428511
H         0.228294     0.360087    -1.453596
H        -0.968541     1.633938    -0.148796
O        -1.028009     1.551241    -1.128689
H        -0.783359     2.439917    -1.455647
O        -0.105925     4.252802    -1.452838
H        -0.357505     4.505249    -0.527236
H        -0.518856     4.895007    -2.043201
O         2.601221     3.656943    -1.488481
H         1.681480     3.998261    -1.523802
H         3.190161     4.419966    -1.487172
H         0.794522    -1.026929    -0.942439
C        -4.209373    -2.238647     0.799422
N        -3.046041    -2.778905     0.447196
C        -1.970606    -3.217209     0.175456
```
==========================================
HOCH2NC-16H2O-H3O+ (intermediate)
==========================================

```
  59
EB3LYP = -1508.155186   ZPE = 309.829
  O    -3.106667   -1.554092    3.001857
  H    -4.070445   -2.186246    1.703662
  H    -3.460442   -1.250409    3.845880
  O    -4.472863   -2.497238    0.856446
  H    -4.666087   -1.441425   -0.119154
  H    -3.838562   -3.130153    0.469352
  O    -2.357771   -3.732346   -0.663137
  H    -1.552559   -3.537628   -0.131098
  H    -2.312627   -4.666536   -0.903691
  O    -0.269297   -2.866503    0.913585
  H    -0.208123   -3.294556    1.778376
  H     0.644696   -2.885046    0.535132
  O     5.703785   -2.036862   -0.132896
  H     5.450100   -0.990968    1.522889
  H     5.979700   -2.957519   -0.024561
  O     3.536412    1.175607   -1.593424
  H     4.614407   -0.886775   -1.412401
  H     4.126344    1.813625   -2.013660
  O     3.536523    1.530376    1.029879
  H     2.624840    1.565600    1.370398
  H     3.609715    1.336840   -0.609343
  O     0.901744    1.889277   -1.551901
  H     1.814153    1.570058   -1.752124
  H     0.249597    1.358979   -2.049545
  O     0.213854    4.537401   -1.284051
  H     0.559214    3.649256   -1.524782
  H     0.733047    5.190478   -1.766464
  O    -1.471228    0.697930   -2.589785
  H    -1.777032    1.232889   -3.334474
  H    -2.002941    0.984727   -1.809055
  O    -2.897877    1.053969   -0.249441
  H    -2.304735    0.579228    0.385734
  H    -3.058821    1.948487    0.153005
  O    -4.673789   -0.721101   -0.892198
  H    -4.114641    0.062730   -0.577479
  H    -4.109021   -1.132356   -1.645315
  H    -2.322624   -1.083059   -2.723766
  O    -3.025499   -1.743018   -2.569687
  H    -2.630211   -2.490601   -2.078308
  H     4.038579    0.881333    1.564619
  O     5.081972   -0.436284    2.231599
  H     5.821875   -0.196663    2.802560
  H    -2.446717   -0.895769    2.704080
  O    -1.213061   -0.220755    1.485234
  H    -0.486992    0.449136    1.534482
  H     0.716963    1.771037    0.239294
  O     0.661493    1.741541    1.222070
  H     0.312457    2.621571    1.466301
  O    -0.542060    4.346947    1.272534
  H    -0.216262    4.585833    0.366907
  H    -0.284771    5.063616    1.865428
  O    -3.131622    3.402147    0.992325
  H    -2.273850    3.859575    1.129968
  H    -3.797757    4.084040    0.849222
  H    -0.799548   -1.069190    1.229874
```

```
C      4.721284    -1.943665    -1.136753
N      3.433690    -2.386342    -0.657806
C      2.385188    -2.689597    -0.238012
H      4.972028    -2.546143    -2.020785
```
============================================
TS (isomerization transition state)
============================================
```
  59
EB3LYP = -1508.094573  ZPE = 308.282
 O     -0.781938    -2.604285     2.899528
 H     -1.986422    -3.514691     1.861049
 H     -0.781604    -2.724462     3.857285
 O     -2.510899    -3.912833     1.132970
 H     -3.449693    -3.006252     0.384618
 H     -1.853837    -4.227683     0.482687
 O     -0.603691    -4.031949    -1.003548
 H      0.057173    -3.455441    -0.553264
 H     -0.120838    -4.785008    -1.366998
 O      0.889416    -2.287838     0.503132
 H      0.867795    -2.629454     1.412414
 H      1.828761    -2.146348     0.261186
 O      5.666003    -0.188025    -0.182489
 H      5.001614     0.623419     1.672751
 H      6.507520    -0.665244    -0.256697
 O      2.608329     1.938577    -1.673167
 H      3.901532     0.017076    -1.157031
 H      3.073893     2.583020    -2.220767
 O      2.665313     2.694900     0.865621
 H      1.756129     2.648600     1.214004
 H      2.697711     2.262151    -0.731415
 O     -0.119373     2.186089    -1.641853
 H      0.837915     2.041517    -1.833400
 H     -0.638264     1.450297    -2.021651
 O     -1.718938     4.425336    -1.511576
 H     -1.049746     3.737779    -1.718030
 H     -1.525633     5.191313    -2.063828
 O     -2.118171     0.292815    -2.388370
 H     -2.683155     0.732070    -3.038121
 H     -2.602507     0.333144    -1.529081
 O     -3.204138     0.036866     0.141635
 H     -2.370427    -0.063548     0.667203
 H     -3.689822     0.805856     0.541429
 O     -3.974037    -2.395951    -0.272432
 H     -3.817305    -1.430119    -0.006108
 H     -3.478027    -2.491196    -1.178270
 H     -2.239486    -1.636180    -2.581076
 O     -2.543820    -2.540421    -2.369167
 H     -1.758054    -3.060229    -2.101102
 H      3.239444     2.183274     1.469405
 O      4.403053     1.046828     2.305681
 H      4.957251     1.364889     3.029006
 H     -0.891165    -1.645395     2.726567
 O     -0.846160    -0.263433     1.471271
 H     -0.541516     0.677497     1.507599
 H     -0.176620     2.296386     0.143030
 O     -0.149802     2.331842     1.127492
 H     -0.816594     3.005773     1.364286
```

```
   O       -2.321118       4.195637       1.096518
   H       -2.142876       4.441946       0.153487
   H       -2.360155       5.018257       1.599789
   O       -4.253787       2.209666       1.271315
   H       -3.677234       3.002752       1.231574
   H       -5.167101       2.517014       1.285513
   H       -0.178338      -0.751239       0.949619
   C        4.810034      -0.585699      -1.139592
   H        5.223432      -0.810224      -2.123638
   C        4.588909      -2.510209      -0.887470
   N        3.566095      -2.055202      -0.478875
==========================================
HOCH2CN-16H2O-H3O+ (glycolonotrile product)
==========================================
  59
EB3LYP = -1508.183422   ZPE = 309.914
   O       -3.064471      -1.546420       3.011652
   H       -4.053414      -2.158329       1.722286
   H       -3.403730      -1.247769       3.863406
   O       -4.471052      -2.459012       0.878751
   H       -4.664746      -1.395288      -0.088253
   H       -3.850082      -3.097502       0.479416
   O       -2.390849      -3.709089      -0.675491
   H       -1.577083      -3.524702      -0.153356
   H       -2.356565      -4.641920      -0.923077
   O       -0.273857      -2.866463       0.879468
   H       -0.204949      -3.298456       1.741547
   H        0.628631      -2.896999       0.487491
   O        5.642442      -2.110960      -0.114078
   H        5.469171      -1.017634       1.507598
   H        5.877114      -3.040608       0.010742
   O        3.546683       1.133740      -1.625601
   H        4.575946      -0.931165      -1.405505
   H        4.138899       1.759659      -2.060556
   O        3.572984       1.514039       0.991567
   H        2.665008       1.552413       1.341847
   H        3.630339       1.308117      -0.644264
   O        0.918294       1.884319      -1.564559
   H        1.823362       1.551210      -1.772953
   H        0.253594       1.363387      -2.055722
   O        0.262830       4.540322      -1.285803
   H        0.597180       3.649417      -1.531400
   H        0.784228       5.189277      -1.771398
   O       -1.477672       0.726964      -2.582591
   H       -1.785277       1.269239      -3.321220
   H       -1.998895       1.014504      -1.795058
   O       -2.875361       1.085451      -0.225739
   H       -2.279087       0.603020       0.400586
   H       -3.023665       1.980014       0.181457
   O       -4.674543      -0.670481      -0.857006
   H       -4.104814       0.106498      -0.544503
   H       -4.122506      -1.082560      -1.619138
   H       -2.348055      -1.045533      -2.718176
   O       -3.056896      -1.698163      -2.560416
   H       -2.664564      -2.454344      -2.080009
   H        4.083427       0.875545       1.531224
   O        5.129016      -0.432873       2.206315
```

```
 H       5.885537    -0.193745     2.755199
 H      -2.403641    -0.890496     2.710502
 O      -1.180892    -0.213541     1.482019
 H      -0.449516     0.450685     1.528875
 H       0.749819     1.767742     0.228620
 O       0.704237     1.737579     1.211838
 H       0.364479     2.619956     1.460706
 O      -0.473907     4.353894     1.277111
 H      -0.153769     4.590688     0.368965
 H      -0.203597     5.066856     1.868682
 O      -3.074258     3.432135     1.024454
 H      -2.211311     3.882421     1.152874
 H      -3.736984     4.119531     0.892308
 H      -0.776699    -1.062191     1.213319
 C       4.654860    -1.993465    -1.131915
 C       3.322484    -2.426074    -0.662100
 N       2.280532    -2.752211    -0.277398
 H       4.916592    -2.569759    -2.032330
==========================================
H2O (desorbed)
==========================================
    3
EB3LYP =   -76.443378   ZPE =   13.338
 H       0.000000     0.765622    -0.469661
 O       0.000000     0.000000     0.117415
 H       0.000000    -0.765622    -0.469661
==========================================
HOCH2CN (desorbed)
==========================================
    7
EB3LYP =  -207.990067   ZPE =   31.611
 O      -1.517877    -0.450853    -0.108292
 H      -1.394615    -1.095720     0.600629
 H      -0.732901     1.284224    -0.798059
 C      -0.569149     0.592304     0.038813
 C       0.827257     0.113261    -0.006613
 N       1.918730    -0.281544    -0.015191
 H      -0.709230     1.155737     0.976898
==========================================
15H2O-H3O+ (ice remnant after desorption)
==========================================
   49
EB3LYP = -1223.720690   ZPE =  261.301
 O       2.710835     0.175477     3.059812
 H       3.730052    -0.316022     1.767668
 H       2.781872    -0.087972     3.984489
 O       4.160962    -0.544094     0.907193
 H       3.529606    -1.608442     0.176755
 H       4.134959     0.263978     0.361197
 O       3.517439     1.486990    -1.061774
 H       2.761320     2.012405    -0.719578
 H       4.156075     2.119250    -1.416081
 O       1.269073     2.648327     0.112436
 H       1.460298     3.328184     0.772814
 H       0.486737     2.980466    -0.390606
 O      -1.111252     3.361185    -1.081095
 H      -1.220998     4.130906    -1.653759
```

```
O   -2.383295    3.347526    1.255328
H   -2.565785    2.402820    1.446513
H   -1.670314    3.514860   -0.273737
O   -2.083850    0.711974   -1.543864
H   -1.805886    1.641712   -1.650322
H   -1.375685    0.152535   -1.931810
O   -4.263807   -1.016812   -1.453200
H   -3.668361   -0.271631   -1.666277
H   -5.047141   -0.929907   -2.008322
O   -0.143364   -1.103919   -2.384410
H   -0.511075   -1.763016   -2.988202
H   -0.003572   -1.559625   -1.517147
O    0.478885   -2.029136    0.114103
H    0.457334   -1.201247    0.655255
H   -0.200158   -2.630081    0.514487
O    3.022915   -2.283013   -0.467742
H    2.073330   -2.359041   -0.146557
H    2.949547   -1.777833   -1.361621
H    1.751581   -0.763010   -2.721193
O    2.708338   -0.800841   -2.530961
H    2.971529    0.089030   -2.221453
H   -3.099036    3.866573    1.639818
H    1.765845    0.154471    2.812400
O    0.391169    0.313455    1.540644
H   -0.582902    0.422258    1.633675
H   -2.269591    0.567816    0.235914
O   -2.330138    0.597193    1.218783
H   -2.918538   -0.149186    1.448231
O   -3.861914   -1.788898    1.104215
H   -4.157246   -1.590623    0.181427
H   -4.654516   -2.008963    1.609461
O   -1.557883   -3.383745    1.194371
H   -2.433800   -2.945145    1.175470
H   -1.707018   -4.326249    1.329029
H    0.688617    1.107389    1.051337
```

**END OF LINE**